\documentclass[conference]{IEEEtran}
\IEEEoverridecommandlockouts
\usepackage{cite}
\usepackage{amsmath,amssymb,amsfonts}
\usepackage{algorithmic}
\usepackage{balance}
\usepackage{graphicx}
\usepackage{textcomp}
\usepackage{xcolor}
\def\BibTeX{{\rm B\kern-.05em{\sc i\kern-.025em b}\kern-.08em
    T\kern-.1667em\lower.7ex\hbox{E}\kern-.125emX}}
\usepackage{changes}
\usepackage{float}
\definecolor{anotherRed}{RGB}{244,195,184}
\definecolor{phase5}{HTML}{EBFCFB}

\usepackage{amsmath}
\usepackage{blkarray}
\usepackage{mathtools, cuted}
\usepackage{lipsum, color}
\usepackage{graphicx}
\usepackage{textcomp}
\usepackage{url}
\usepackage[utf8]{inputenc}
\usepackage[T1]{fontenc}
\usepackage{array}
\newcolumntype{P}[1]{>{\centering\arraybackslash}p{#1}}
\usepackage[nameinlink]{cleveref}
\usepackage{comment}
\usepackage{multirow}
\usepackage{multicol}

\usepackage{xcolor}
\usepackage{color,soul}
\usepackage{fontawesome}
\usepackage{pifont}
\usepackage{blkarray}
\usepackage{tikz}
\usetikzlibrary{arrows,automata,arrows.meta}

\definecolor{anotherRed}{RGB}{244,195,184}
\definecolor{phase5}{HTML}{EBFCFB}

\newcommand*{\twoelementtable}[3][l]%
{%  
    \begin{tabular}[t]{@{}#1@{}}%
        #2\tabularnewline
        #3%
    \end{tabular}%
}

\usepackage{bbding}

\usepackage{csquotes}
\usepackage{amsmath}

\usepackage{longtable}
\usepackage[acronym,toc=true]{glossaries}
\usepackage{glossary-longragged} 

\newacronym{ALS}{ALS}{Adaptive learning systems}
\newacronym{APG}{APG}{Automatic Problem Generation}
\newacronym{CTF}{CTF}{Capture the Flag}
\newacronym{EEA}{EEA}{Electronic Exercise Assistant}
\newacronym{ITS}{ITS}{intelligent tutoring system}
\newacronym{FBS}{FBS}{Flying Base Station}
\newacronym{AI}{AI}{Artificial Intelligence}
\newacronym{API}{API}{Application Programming Interface}
\newacronym{CoAP}{CoAP}{Constrained Application Protocol}
\newacronym{CLP}{CLP}{Covering Location Problem}
\newacronym{BTS}{BTS}{Base Tranceiver Station}
\newacronym{BFS}{BFS}{Breadth-first Search}
\newacronym{CSS}{CSS}{Cascading Style Sheets}
\newacronym{CRUD}{CRUD}{Create Read Update Delete}
\newacronym{CUCKS}{CUCKS}{Cuckoo Search}
\newacronym{GA}{GA}{Genetic Algorithm}
\newacronym{HTTP}{HTTP}{Hyper Text Transfer Protocol}
\newacronym{HTML}{HTML}{HyperText Markup Language}
\newacronym{IaaS}{IaaS}{Infrastructure as a Service}
\newacronym{IBM}{IBM}{International Business Machines}
\newacronym{ID}{ID}{Identifier}
\newacronym{IDE}{IDE}{Integrated Development Environment}
\newacronym{IoT}{IoT}{Internet of Things}
\newacronym{IT}{IT}{Information Technology}
\newacronym{IP}{IP}{Internet Protocol}
\newacronym{IPsec}{IPsec}{Internet Protocol Security}
\newacronym{LTE}{LTE}{Long Term Evolution}
\newacronym{LP}{LP}{Linear Programming}
\newacronym{JPA}{JPA}{Java Persistence API}
\newacronym{JSP}{JSP}{Java Server Pages}
\newacronym{JSON}{JSON}{JavaScript Object Notation}
\newacronym{MQTT}{MQTT}{MQ Telemetry Transport}
\newacronym{NIST}{NIST}{National Institute of Standards and Technology}
\newacronym{NoSQL}{NoSQL}{No Structure Query Language}
\newacronym{ORM}{ORM}{Object Relation Mapping}
\newacronym{OS}{OS}{Operating System}
\newacronym{PaaS}{PaaS}{Platform as a Service}
\newacronym{REST}{REST}{Representational State Transfer}
\newacronym{REST API}{REST API}{Representational State Transfer Application Programming Interface}
\newacronym{RFID}{RFID}{Radio Frequency Identification}
\newacronym{ROI}{ROI}{Return on Investment}
\newacronym{RPC}{RPC}{Remote Procedure Call}
\newacronym{RSRP}{RSRP}{Reference Signal Received Power}
\newacronym{RSSI}{RSSI}{Received Signal Strength Indicator}
\newacronym{SA}{SA}{Simulated Annealing}
\newacronym{SaaS}{SaaS}{Software as a Service}
\newacronym{SOAP}{SOAP}{Simple Object Access Protocol}
\newacronym{SLA}{SLA}{Service Level Agreement}
\newacronym{SLAs}{SLAs}{Service Level Agreements}
\newacronym{UDP}{UDP}{User Datagram Protocol}
\newacronym{URI}{URI}{Uniform Resource Identifier}
\newacronym{URL}{URL}{Uniform Resource Locator}
\newacronym{XML}{XML}{Extensible Markup Language}
\newacronym{MTCGs}{MTCGs}{Machine-type Communication Gateways}
\newacronym{LPWANs}{LPWANs}{Low-Power Wide-Area Networks}
\newacronym{MCS}{MCS}{Modulation and Coding Scheme}
\newacronym{CIoT}{CIoT}{Cellular IoT}
\newacronym{NB-IoT}{NB-IoT}{Narrowband IoT}
\newacronym{LPWA}{LPWA}{Low-Power Wide-Area}
\newacronym{WSDL}{WSDL}{Web Services Description Language}
\newacronym{SMTP}{SMTP}{Simple Mail Transfer Protocol}
\newacronym{UAV}{UAV}{Unmanned Aerial Vehicle}
\newacronym{KSP}{KSP}{Knapsack Problem}
\newacronym{WANET}{WANET}{Wireless Ad Hoc Network}
\newacronym{RSRPC}{RSRPC}{Reference Signal Received Power Categories}
\newacronym{SCP}{SCP}{Set Covering Problem}
\newacronym{MCP}{MCP}{Maximum Coverage Problem}
\newacronym{LSCP}{LSCP}{Location Set Covering Problem}
\newacronym{MS-LSCP}{MS-LSCP}{Multi-Service Location Set Covering Problem}
\newacronym{MCLP}{MLCP}{Maximal Covering Location Problem}
\newacronym{ESS-LSCP}{ESS-LSCP}{Existing Service System LSCP}
\newacronym{FT-LSCP}{FT-LSCP}{Facility Type LSCP}
\newacronym{MO-LSCP}{MO-LSCP}{Multi Objectives LSCP}
\newacronym{MEXCLP}{MEXCLP}{Maximal Expected Covering Problem}
\newacronym{SQ-MCLP}{SQ-MCLP}{Site Quality MCLP}
\newacronym{CLSCP}{CLSCP}{Capacity Location Set Covering Problem}
\newacronym{CLSCP-SO}{CLSCP-SO}{Capacitated LSCP -- System Optimal}
\newacronym{MS-CLSCP}{MS-CLSCP}{Multi-Service Capacitated Location Set Covering Problem}
\newacronym{MALP}{MALP}{Maximal Availability Location Problem}
\newacronym{CLSCP-CA}{CLSCP-CA}{Capacitated LSCP with Closest Assignment}
\newacronym{GMCLP}{GMCLP}{Generalized MCLP}
\newacronym{CLSCP-EA}{CLSCP-EA}{Capacitated LSCP with Equal Assignment}
\newacronym{LSCP-FT}{LSCP-FT}{Location Set Covering Problem with Facility Types}

\newacronym{HS}{HS}{Harmony Search}
\newacronym{DE}{DE}{Differential Evolution}
\newacronym{SOA}{SOA}{Seeker Optimization Algorithm}
\newacronym{EP}{EP}{Evolutionary Programming}
\newacronym{ACO}{ACO}{Ant Colonies Optimization}
\newacronym{PSO}{PSO}{Particle Swarm Optimization}
\newacronym{ABC}{ABC}{Artificial Bee Colony}
\newacronym{GSA}{GSA}{Gravitational Search Algorithm}
\newacronym{FA}{FA}{Firefly Algorithm}
\newacronym{TLA}{TLA}{Teaching Learning Algorithm}
\newacronym{CRO}{CRO}{Chemical Reaction Optimization}
\newacronym{WCA}{WCA}{Water Cycle Algorithm}
\newacronym{DSA}{DSA}{Differential Search Algorithm}
\newacronym{MOA}{MOA}{Meta-heuristic Optimization Algorithms}

\newacronym{GAMS}{GAMS}{General Algebraic Modeling System}

\newacronym{3NF}{3NF}{Third Normal Form}
\newacronym{ACID}{ACID}{Atomicity Consistency Isolation Durability}
\newacronym{BASE}{BASE}{Basically Available Soft-State Eventual consistency}
\newacronym{CAP}{CAP}{Consistency Availability Partition tolerance}
\newacronym{CMN}{CMN}{Cloud Management Node}
\newacronym{DAO}{DAO}{Database Access Object}
\newacronym{DTO}{DTO}{Data Transfer Object}
\newacronym{ES}{ES}{Elasticsearch}
\newacronym{FTE}{FTE}{Full Time Employee}
\newacronym{HATEOAS}{HATEOAS}{Hypermedia as the Engine of Application State}
\newacronym{Java EE}{Java EE}{Java Enterprise Edition}
\newacronym{JVM}{JVM}{Java Virtual Machine}
\newacronym{KYPO}{KYPO}{Kybernetický polygon}
\newacronym{LMN}{LMN}{Lan Management Node}
\newacronym{MVC}{MVC}{Model View Controller}
\newacronym{ODM}{ODM}{Object Document Mapping}
\newacronym{OLAP}{OLAP}{Online Analytical Processing}
\newacronym{OLTP}{OLTP}{Online Transaction Processing}
\newacronym{OSDI}{OSDI}{Operating Systems Design and Implementation}
\newacronym{POJO}{POJO}{Plain Old Java Object}
\newacronym{SLF4J}{SLF4J}{Simple Logging Facade for Java}
\newacronym{SMN}{SMN}{Sandbox Management Node}
\newacronym{SQL}{SQL}{Structured Query Language}
\newacronym{SSH}{SSH}{Secure Shell}
\newacronym{URN}{URN}{Uniform Resource Name}
\newacronym{UUID}{UUID}{Universally Unique Identifier}
\newacronym{VDS}{VDS}{Visualization Data Service}
\newacronym{VNC}{VNC}{Virtual Network Computing}
\newacronym{WBS}{WBS}{Work Breakdown Structure}

\let\oldgls\gls
\renewcommand{\gls}{\oldgls*}

\newlength{\acronymlabelwidth}
\newglossarystyle{listwithwidth}{%
    {\begin{description}}{\end{description}}%
  \renewcommand*{\glsgroupheading}[1]{}%
}

\usepackage{pgfplots}
\pgfplotsset{compat=1.8}
\usepgfplotslibrary{statistics}
\usepackage{cleveref}

\pgfplotsset{
    box plot/.style={
        /pgfplots/.cd,
        black,
        only marks,
        mark=-,
        font=\sffamily,
        mark size=1em,
        /pgfplots/error bars/.cd,
        y dir=plus,
        y explicit,
    },
    box plot box/.style={
        /pgfplots/error bars/draw error bar/.code 2 args={%
            \draw  ##1 -- ++(1em,0pt) |- ##2 -- ++(-1em,0pt) |- ##1 -- cycle;
        },
        /pgfplots/table/.cd,
        y index=2,
        y error expr={\thisrowno{3}-\thisrowno{2}},
        /pgfplots/box plot
    },
    box plot top whisker/.style={
        /pgfplots/error bars/draw error bar/.code 2 args={%
            \pgfkeysgetvalue{/pgfplots/error bars/error mark}%
            {\pgfplotserrorbarsmark}%
            \pgfkeysgetvalue{/pgfplots/error bars/error mark options}%
            {\pgfplotserrorbarsmarkopts}%
            \path ##1 -- ##2;
        },
        /pgfplots/table/.cd,
        y index=4,
        y error expr={\thisrowno{2}-\thisrowno{4}},
        /pgfplots/box plot
    },
    box plot bottom whisker/.style={
        /pgfplots/error bars/draw error bar/.code 2 args={%
            \pgfkeysgetvalue{/pgfplots/error bars/error mark}%
            {\pgfplotserrorbarsmark}%
            \pgfkeysgetvalue{/pgfplots/error bars/error mark options}%
            {\pgfplotserrorbarsmarkopts}%
            \path ##1 -- ##2;
        },
        /pgfplots/table/.cd,
        y index=5,
        y error expr={\thisrowno{3}-\thisrowno{5}},
        /pgfplots/box plot
    },
    box plot median/.style={
        /pgfplots/box plot,
    },
}

\newcommand\copyrighttext{%
  \footnotesize \textcopyright 2021 IEEE. Personal use of this material is permitted. Permission from IEEE must be obtained for all other uses, in any current or future media, including reprinting/republishing this material for advertising or promotional purposes, creating new collective works, for resale or redistribution to servers or lists, or reuse of any copyrighted component of this work in other works. Cite this article as follows: P. Seda, J. Vykopal, V. Švábenský, and P. Čeleda \textit{Reinforcing Cybersecurity Hands-on Training With Adaptive Learning}, in Proceedings of the 51st IEEE Frontiers in Education Conference (FIE '21). Lincoln, Nebraska, USA, 2021. DOI: 10.1109/FIE49875.2021.9637252.}
\newcommand\copyrightnotice{%
\begin{tikzpicture}[remember picture,overlay]
\node[anchor=south,yshift=12pt] at (current page.south) {\fbox{\parbox{\dimexpr\textwidth-\fboxsep-\fboxrule\relax}{\copyrighttext}}};
\end{tikzpicture}%
}

\begin{document}
%\sloppy

\title{Reinforcing Cybersecurity Hands-on Training \\ With Adaptive Learning}

\author{\IEEEauthorblockN{Pavel Seda}
\IEEEauthorblockA{%Masaryk University, Faculty of Informatics, Brno, Czech Republic \\
\textit{Faculty of Informatics} \\
\textit{Masaryk University} \\
Brno, Czech Republic\\
seda@fi.muni.cz
\and
\IEEEauthorblockN{Jan Vykopal}
\IEEEauthorblockA{%Masaryk University, Faculty of Informatics, Brno, Czech Republic \\
\textit{Faculty of Informatics} \\
\textit{Masaryk University} \\
Brno, Czech Republic\\
vykopal@fi.muni.cz
}
\and
\IEEEauthorblockN{Valdemar Švábenský}
\IEEEauthorblockA{%Masaryk University, Faculty of Informatics, Brno, Czech Republic \\
\textit{Faculty of Informatics} \\
\textit{Masaryk University} \\
Brno, Czech Republic\\
svabensky@fi.muni.cz
}
\and
\IEEEauthorblockN{Pavel Čeleda}
\IEEEauthorblockA{%Masaryk University, Faculty of Informatics, Brno, Czech Republic \\
\textit{Faculty of Informatics} \\
\textit{Masaryk University} \\
Brno, Czech Republic\\
celeda@fi.muni.cz
}
}}
%%%%%%%%%%%%%%%%%%%%%%%%%%%%%%%%%%%%%%%%%%%%%%%%%%%%%%%%%%%%%%%%%%%%%%%%%%%%%%%%%%%%
%%%%%%%%%%%%%%%%%%%%% END COMMENTED AUTHORS LIST %%%%%%%%%%%%%%%%%%%%%%%%%%%%%%%%%%%
%%%%%%%%%%%%%%%%%%%%%%%%%%%%%%%%%%%%%%%%%%%%%%%%%%%%%%%%%%%%%%%%%%%%%%%%%%%%%%%%%%%%

\maketitle
\copyrightnotice

\begin{abstract}
This Research To Practice Full Paper presents how learning experience influences students' capability to learn and their motivation for further learning. Although each student is different, standard instruction methods do not adapt to individual students. Adaptive learning reverses this practice and attempts to improve the student experience. While adaptive learning is well-established in programming, it is rarely used in cybersecurity education. This paper is one of the first works investigating adaptive learning in cybersecurity training. First, we analyze the performance of 95 students in 12 training sessions to understand the limitations of the current training practice. 
Less than half of the students (45 out of 95) completed the training without displaying any solution, and only in two sessions, all students completed all phases. 
Then, we simulate how students would proceed in one of the past training sessions if it would offer more paths of various difficulty.
Based on this simulation, we propose a novel tutor model for adaptive training, which considers students' proficiency before and during an ongoing training session. The proficiency is assessed using a pre-training questionnaire and various in-training metrics. 
Finally, we conduct a case study with 24 students and new training using the proposed tutor model and adaptive training format. The results show that the adaptive training does not overwhelm students as the original static training format. In particular, adaptive training enables students to enter several alternative training phases with lower difficulty than the phases in the original training. The proposed adaptive format is not restricted to particular training used in our case study. Therefore, it can be applied to practicing any cybersecurity topic or even in other related computing fields, such as networking or operating systems. Our study indicates that adaptive learning is a promising approach for improving the student experience in cybersecurity education. We also highlight diverse implications for educational practice that improve students' experience.
\end{abstract}


\section{Introduction}
\label{sec:intro}

Learning cybersecurity requires extensive knowledge and skills, ranging from a wide area of theoretical concepts to practical skills with operating systems, command-line tools, and system vulnerabilities~\cite{mouheb2019cybersecurity}. As~a~result, it is difficult to conduct hands-on cybersecurity training that would match the skills of all participants in the training. This situation is further complicated since more and more students with different backgrounds are entering the field of cybersecurity~\cite{bashir2017profiling}.

Although the instructor can intervene to help students interactively, this is feasible only in relatively small classes, and not every student actively asks for help. The interactive help is especially complicated during online training (e.g., forced by restrictions caused by the COVID-19 pandemic \cite{putri2020impact}).

To support our assumptions that students do not fully benefit from the training sessions, we analyze 12 hands-on training sessions on various cybersecurity topics we held in 2019 and 2020. We observed that only 47\% of students successfully completed the training~(for more information, see \Cref{sec:motivation}).

We see the opportunity to improve the students' experience and skills using an \gls{ITS}, which adapts the learning environment according to the student's abilities. Unfortunately, an ITS in the domain of hands-on cybersecurity training is rare, mostly because the interactive lab environment and its setup differ for particular sessions. 
As a result, cybersecurity platforms offer static scenarios with limited or no adaptiveness \cite{braghin2020modeldriven}. We could create an ITS for a specific training session. This would bring great flexibility in defining the conditions for serving adaptive tasks to students. However, such ITS could not be reused for another training. Our main goal is to create a concept of generic cybersecurity training that will adapt to the current phase of individual student skills.

In this paper, we present a generic format for adaptive training and a tutor model. The model determines appropriate tasks based on students' theoretical knowledge and current performance. Using the proposed format and model, we conduct a case study involving cybersecurity hands-on training with 24 undergraduate students and graduates in computer science. We report teaching experience from the execution of adaptive hands-on training based on the proposed tutor model implemented in KYPO Cyber Range Platform (CRP)~\cite{2021-FIE-kypo-csc}. The results suggest that adaptive training increases the chances of successful completion of training and deepens the experience and knowledge gained from the training. In our study, 88\% of students completed the training without asking for a solution of any task. Further, most of the students reported that they did not get stuck at any point of the training and enjoyed it. Finally, we provide recommendations for instructors on using the proposed format and model and also depict future research directions.

This paper is organized into six sections. \Cref{sec:related_work} provides an overview of \gls{ITS}s in computer science education. \Cref{sec:previousTeachingExperience} describes our past experience and motivation. \Cref{sec:design} details the training format and the newly developed tutor model. \Cref{sec:case_study_setup} describes case study setup, including teaching context and participants. \Cref{sec:results} reports the results from three hands-on training sessions. Finally, \Cref{sec:conclusion} concludes the paper and outlines future work.

\section{Related Work}
\label{sec:related_work}

Adaptive learning techniques are a well-established research area~\cite{colchester2017survey} that accommodates the pedagogical content for the learners and their current state of knowledge. These techniques were introduced in the 1970s \cite{carbonell1970ai}, and the research area still receives considerable interest. Personalized learning achievable by adaptive techniques was identified by the US National Academy of Engineering as one of the Grand Challenges for Engineering~\cite{grand_challenges}.

We start with \gls{ITS} that conceptualize adaptive learning in a way that is commonly accepted in computer science. \gls{ITS} typically contain the following parts: (i) \textit{domain model}, (ii) \textit{student model}, (iii) \textit{tutor model}, and (iv) \textit{user interface model}~\cite{sottilare2016design}. The \textit{domain model} presents educational content and its relationships~\cite{sottilare2016design}. The \textit{student model} captures the students' knowledge to assess their performance~\cite{long2011students,hu2014content}. The \textit{tutor model} (\textit{instructional policy}) presents the suitable learning tasks to students~\cite{nkambou2010advances}. Finally, a \textit{user interface model} interacts with the user via a pre-defined interface \cite[chapter 9]{sottilare2016design}. 

Although the \gls{ITS} research area is well established, to the best of our knowledge, there are no available \gls{ITS} models for comprehensive hands-on cybersecurity training in a networked lab environment. For that reason, we review the \gls{ITS} research from other domains, which improve or discuss student models to evaluate the participants' performance and tutor models to assign suitable tasks. Effenberger and Pelánek \cite{effenberger2019measuring} discuss several approaches to measure the student's performance during introductory programming tasks. They find that the widely used performance measure called binary success is not suitable for the evaluation of programming tasks since it contains too little information. The evaluation of programming tasks is harder than the evaluation of answering multiple-choice questions about any topic in computer science. Therefore, they propose multiple qualitative and quantitative methods, based on the four performance levels \textit{failed}, \textit{poor}, \textit{good}, and \textit{excellent}. Khosravi et. al \cite{khosravi2020development} provide lessons learned from using the Ripple system that recommends suitable learning activities for students of relational databases. The authors found that an important part of the learning system is based on gamification, such as awards and leaderboards to motivate students. Further, \cite{hatzivasilis2020modern} uses Bloom's taxonomy to dynamically adapt the training process. The authors define several layers with different difficulties that should be accomplished. The system evaluates the students' exercises and exams during the training. After reaching a good understanding, the student can proceed with a related advanced training scenario. Contrary to our approach, it seems that their adaptiveness is mostly based on exam scores and does not include more detailed metrics such as the commands used in an interactive learning environment. For more information on \gls{ITS}, we suggest \cite{sottilare2016design} that focuses on design recommendations and \cite{colchester2017survey,paladines2020systematic,alkhatlan2018intelligent} that review the recent research.

Next, the participants' perceptions of difficulty are subjective. Nebel~et.~al~\cite{nebel2020competitive} discussed that perceived difficulty within a competition might differ relative to each learner's performance. A participant winning effortlessly might indicate a low difficulty, whereas a losing participant may perceive a relatively high difficulty even if the context is identical. This argumentation appears evident but is important. The individual difficulty might play a crucial role in influencing the students' experience and how the learning process evolves. Xue~et~al.~\cite{xue2017dynamic} observed that perceiving the difficulty influences participant engagement and how often the training is played.

Finally, we searched for related works in the area of cybersecurity education. We found only a few relevant sources about adaptiveness in cybersecurity training. Hatzivasilis~et~al.~\cite{hatzivasilis2020modern} propose suitable assignments of cybersecurity tasks to students in exercises held in a cyber range. However, they do not propose the unified design of adaptive hands-on training. In the industry sector, the Circadence company provides adaptable cyber training and learning opportunities. However, their platform does not support adaptive task assignments based on the students' performance and mainly focuses on the adaptive pre-configuration of training sessions. This includes turning the hints and chatbot on or off during the training~\cite{circadance2021}. 

Based on the available literature and eight years of our experience with hands-on cybersecurity training, we believe the reason for the absence of ITS in comprehensive hands-on cybersecurity education is the high complexity of systems (hardware, software, and domain knowledge requirements).

\section{Our Teaching Experience and Expectations from Adaptive Learning}\label{sec:previousTeachingExperience}

This section presents our previous experience with non-adaptive training sessions and our expectations from introducing adaptivity to hands-on training.

\subsection{Our Teaching Experience}\label{sec:motivation}
We have been designing and organizing cybersecurity training sessions since 2014 \cite{198087}. The participating high-school and undergraduate students, as well as professional learners, value the hands-on nature of these sessions and the opportunities to practice cyber attacks and defense. On the other hand, many participants were frustrated in various phases of the training, even though it contained on-demand hints. The participants lacked some prerequisite skills and knowledge or wanted to complete the training without help.

To validate our assumptions about the factors influencing students' learning experience, we analyzed interaction data from 12 training sessions held in 2019 and 2020. The data were collected automatically in the KYPO CRP~\cite{2021-FIE-kypo-csc}. A total of 95 students participated in one of 12 cybersecurity training sessions. Each training comprised three to six consecutive phases. In total, less than half of the participants (45 out of 95) completed their training sessions, i.e., completed all phases without displaying any solution. In two training sessions, all participants completed all phases. In the other ten sessions, the ratio of successful participants ranged from 0 to 83\% (median 55\%). The count of phases that participants completed in the same training session varied too.

These student difficulties can be mitigated by conducting training sessions that adapt to the proficiency and current progress of each student. However, conducting such adaptive training sessions is infeasible without a training tutor integrated into the platform. To support this argument, we counted the actions the students performed in the previous training sessions (see~\Cref{tab:summarizedResultsFromPastNonAdaptiveGames}). These actions include starting the training phase, submitting the correct or incorrect answer in a phase, and displaying a hint or solution. All these actions are automatically processed by the tutor without instructor intervention. In the analyzed training sessions, the average number of actions per participant ranged from 17 to 62 (median 29). This number is too high to conduct the training sessions manually (by the instructor) without the support of the software in the learning environment.

\begin{table}[h]
    \centering
    \caption{Statistics of the past non-adaptive training sessions.}
    \label{tab:summarizedResultsFromPastNonAdaptiveGames}
    \begin{tabular}{P{1.0cm}|P{2.1cm}|P{2.2cm}|P{1.85cm}}
    \hline
    \textbf{Training session} & 
    \textbf{Completed by participants [\%]} &
    \textbf{Most participants ended in phase} &
    \textbf{Avg actions per participant}\\ \hline
    1  & 100  & 3 out of 3 & 28 \\
    2  & 100 & 4 out of 4 & 21 \\
    3  & 0   & 2 out of 4 & 29 \\
    4  & 83  & 2 out of 5 & 29 \\
    5  & 60  & 3 out of 5 & 25 \\
    6  & 0   & 3 out of 5 & 40 \\
    7  & 57  & 1 out of 5 & 17 \\
    8  & 25  & 1 out of 5 & 45 \\
    9  & 18  & 3 out of 6 & 62 \\
    10 & 52  & 3 out of 6 & 29 \\
    11 & 33  & 2 out of 5 & 35 \\
    12 & 66  & 5 out of 5 & 30 \\
    \hline
    \end{tabular}
\end{table}

\subsection{Adaptive Learning Expectations}
Our initial assumption for the integration of adaptivity to the training was that fewer students will fail the training. Further, we suppose they finish the training to the best of their capability and thus fully benefit from the training.

Since adaptive learning was not used in the previous cybersecurity hands-on training, we simulated how students would proceed in one of our previous training sessions, which we made adaptive to students' proficiency and performance. 
We chose a training with six phases including (i) network reconnaissance using \texttt{nmap}, (ii) finding a vulnerability, (iii) exploiting the vulnerability using \texttt{Metasploit}, (iv) Linux operations, (v) cracking a SSH passphrase, and (vi) connecting via SSH using the cracked passphrase and displaying the content of the file.

In our simulation, the adaptivity of the training lies in modifying the difficulty of the tasks presented to each student in all six training phases. We created two new tasks for each phase that contains one or more hints in the assignment to simplify the phase. Next, we selected the metrics gathered in the KYPO CRP.
The metrics used for our simulation were: (i) \emph{pre-training assessment}, (ii) \emph{training completion time}, and (iii) \emph{actions in the learning environment including entered commands}. The pre-training assessment is a questionnaire before the training that maps the theoretical knowledge and self-assessment of skills of the participants relevant to the training. The training completion time captures how long the participant solved a training phase. The actions in the learning environment are commands entered in the learning environment during the training, submissions of the wrong answers, or displaying the solution of the task. In particular, we count a number of entered commands relevant to a particular phase. For instance, too many entered \texttt{ssh} commands may indicate that a participant lacks skills in using this particular command. We employed these metrics to find the most suitable task in each phase for the participant, as shown in~\Cref{tab:predicates}.

\begin{table}[h!]
    \centering
    \caption{Metrics used for determining the most suitable task \\(\faCheck\ = metric used, \faRemove\ = metric not used).}
    \label{tab:predicates}
    \begin{tabular}{P{2.2cm}|P{1.6cm}|P{1.6cm}|P{1.0cm}}
    \hline
        \textbf{Training phase} & \textbf{Pre-training assessment} & \textbf{Performance} & \textbf{Actions} \\  \hline
        1 & \faCheck  & \faRemove & \faRemove \\ 
        2 & \faCheck  & \faCheck  & \faRemove \\
        3 & \faCheck  & \faCheck  & \faCheck  \\
        4 & \faCheck  & \faCheck  & \faRemove \\
        5 & \faCheck  & \faCheck  & \faRemove \\
        6 & \faRemove & \faCheck  & \faCheck \\
    \hline
    \end{tabular}
\end{table}

\begin{figure}[h!]
    \centering
    \includegraphics[clip, trim=2cm 12.40cm 8cm 1.74cm, width=0.5\textwidth]{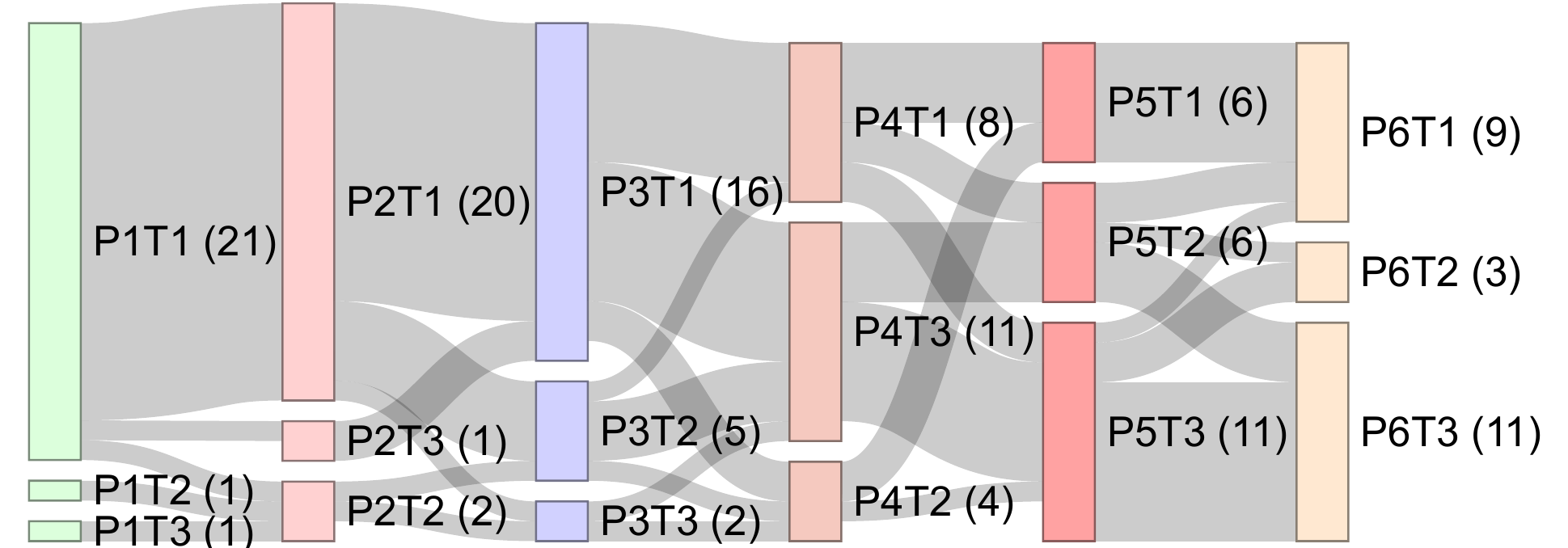}
    \caption{Transitions of participants between particular tasks in training. P\emph{X}T\emph{Y} denotes task T\emph{Y} in the phase P\emph{X}. The number of participants who attempted to solve the task is in brackets.}
    \label{fig:adaptiveLevelsFlow}
\end{figure}

We developed simulation software that processes the data from the non-adaptive training session to calculate the transitions of participants between variant tasks based on the described metrics.
The simulated transitions of 23 participants are shown in a Sankey chart in \Cref{fig:adaptiveLevelsFlow}. The original, non-adaptive training consisted of six tasks: P1T1, P2T1, P3T1, P4T1, P5T1, and P6T1. The newly added, alternative tasks are those denoted T2 or T3, i.e., P1T2, P1T3, P2T2, P2T3, etc.

\begin{figure*}[h]
    \centering
    \begin{tikzpicture}[
        ->,                  % makes the edges directed
        >=stealth',          % makes the arrow heads bold
        shorten >=1pt,    
        auto,
        node distance=1.60cm,  % minimum distance between two nodes
        thick,
        scale=0.50,
        every node/.style={scale=0.700}, % oriģinally scale=0.8
        font=\sffamily
    ]
        \node[state,color=black]   (I)                    {Intro};
        \node[state,color=teal] (A) [right of=I] {A};
        
        \node[state,color=blue] (UD1) [right of=A]       {P$_\text{D}$};
        
        \node[state] (U12) [right of=UD1]       {T$2$};
        \node[state] (U11) [above of=U12]       {T$1$};
        \node[state] (U13) [below of=U12]       {T$3$};
        \node[draw=none] (Unit1) [above of=U11,yshift=-0.60cm] {Phase 1};
        
        \node[state,color=blue] (UD2) [right of=U12]       {P$_\text{D}$};
        
        \node[state] (U22) [right of=UD2]       {T$2$};
        \node[state] (U21) [above of=U22]       {T$1$};
        \node[state] (U23) [below of=U22]       {T$3$};
        \node[draw=none] (Unit2) [above of=U21,yshift=-0.60cm] {Phase 2};
        
        \node[state, color=blue] (UD3) [right of=U22] {P$_\text{D}$};
        
        \node[state] (U32) [right of=UD3]       {T$2$};
        \node[state] (U31) [above of=U32]       {T$1$};
        \node[state] (U33) [below of=U32]       {T$3$};
        \node[draw=none] (Unit3) [above of=U31,yshift=-0.60cm] {Phase 3};
        
        \node[state, color=blue] (UD4) [right of=U32] {P$_\text{D}$};
        
        \node[state] (U42) [right of=UD4]       {T$2$};
        \node[state] (U41) [above of=U42]       {T$1$};
        \node[state] (U43) [below of=U42]       {T$3$};
        \node[draw=none] (Unit4) [above of=U41,yshift=-0.60cm] {Phase 4};

        \node[state, color=blue] (UD5) [right of=U42] {P$_\text{D}$};

        \node[state] (U52) [right of=UD5]       {T$2$};
        \node[state] (U51) [above of=U52]       {T$1$};
        \node[state] (U53) [below of=U52]       {T$3$};
        \node[draw=none] (Unit5) [above of=U51,yshift=-0.60cm] {Phase 5};
        
        \node[state,color=teal] (Q) [right of=U52] {Q};

        \node[state,color=purple] (END) [right of=Q]       {End};
        
        \path (I) edge node {} (A)
        
        (A) edge node {} (UD1)
        
        (UD1) edge node {} (U11)
              edge node {} (U12)
              edge node {} (U13)
        
        (U11) edge node {} (UD2)
        (U12) edge node {} (UD2)
        (U13) edge node {} (UD2)
        
        (UD2) edge node {} (U21)
              edge node {} (U22)
              edge node {} (U23)
        
        (U21) edge node {} (UD3)
        (U22) edge node {} (UD3)
        (U23) edge node {} (UD3)
        
        (UD3) edge node {} (U31)
              edge node {} (U32)
              edge node {} (U33)
              
        (U31) edge node {} (UD4)
        (U32) edge node {} (UD4)
        (U33) edge node {} (UD4)
        
        (UD4) edge node {} (U41)
              edge node {} (U42)
              edge node {} (U43)

        (U41) edge node {} (UD5)
        (U42) edge node {} (UD5)
        (U43) edge node {} (UD5)

        (UD5) edge node {} (U51)
              edge node {} (U52)
              edge node {} (U53)

        (U51) edge node {} (Q)
        (U52) edge node {} (Q)
        (U53) edge node {} (Q)
        
        (Q) edge node {} (END)
        
        ;
    \end{tikzpicture}
    \caption{Graph structure of adaptive cybersecurity training with pre-training assessment (A), decision component (P$_\text{D}$) applying the proposed model, and a~post-training questionnaire (Q). This exemplary training contains five phases. Each phase contains one base task (T1) and two variant tasks (T2, T3).}
    \label{fig:adaptive-game-structure-example}
\end{figure*}
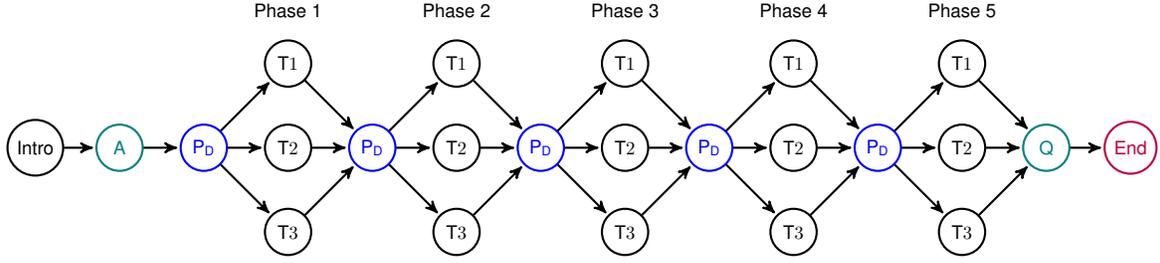

We see the participants would enter not only the original tasks (T1) but also new easier variant tasks (T2 or T3), which indicates the adaptive training would be beneficial for our diversely performing participants. In particular, 17 out of 23 participants would benefit from this adaptive training because they would get one or more variant tasks matching their skills better. These results strengthen our expectation that adaptive learning techniques may increase the students' experience and reduce the number of students that get stuck during the hands-on training.

\begin{align*}
    x= & \text{ the phase a student is entering},\\
    T_{x}= & \text{ the most suitable task of the phase $x$ for the student}, \\
    n_x= & \text{ the number of variant tasks in the phase $x$}, \\
    p_i= & 
    \begin{cases} 
        1, \: \text{if}\: \text{question group $i$ from A is correctly answered}\\
        0, \: \text{otherwise},
    \end{cases} \\
    k_i= & \text{ commands corresponding to the phase $i$ were used}, \\
    e_i= & \text{ expected time to complete of the phase $i$}, \\
    o_i= & \text{ student’s completion time in the phase $i$}, \\
    t_i= &
    \begin{cases} 
        1, \: \text{if}\: o_i < e_i \text{ in phase $i$}\\
        0, \: \text{otherwise}, 
    \end{cases}\\
    s_i= &
    \begin{cases}
         1, \: \text{if}\: \text{ the solution of the phase $i$ is \textit{not} displayed}\\
         0, \: \text{otherwise}, 
    \end{cases}\\
    a_i= & \text{ answers corresponding to the phase $i$ were submitted}.
\end{align*}

Nevertheless, the software was specifically developed for one training and does not provide a generic solution for cybersecurity training with different topics in phases and different relations between its phases. We address this limitation in the next section.

\section{Design of Adaptive Cybersecurity Training} \label{sec:design}
In this section, we present a generic format of adaptive cybersecurity hands-on training based on a model that uses the students' knowledge and performance to assign suitable training tasks.
We evaluate the format using a case study presented in~\Cref{sec:case_study_setup}.

\subsection{Training Format}
We propose a generic structure for adaptive cybersecurity training. \Cref{fig:adaptive-game-structure-example} shows an example of such structure with five phases, each with three tasks of various difficulty. In general, the training can contain an arbitrary number of phases and tasks. The training consists of several components: the introduction (Intro), the pre-training assessment (A), training phases including variant tasks (T\emph{X}), decision components (P$_\text{D}$), and post-training questionnaire (Q).

First, the introduction familiarizes the student with the training and communicates all necessary information before the training start. 

The pre-training assessment is the first component collecting data about students' knowledge and skills. The questions asked in the pre-training assessment are grouped into the \emph{question groups} by their relation to specific training phases. Each question can be assigned into several question groups since they can be relevant to more phases. For each training phase, we set the \textit{essential ratio} of knowledge to determine whether the student's theoretical knowledge or self-reported skills are sufficient or not. For example, the essential ratio can be set to 100\%, which would mean the students need to know the answer to all the questions or self-report a defined level of skills for a particular phase. In particular, pre-training assessment should mostly include knowledge quizzes, as students' self-assessment can be misleading \cite{vsvabensky2018challenges,mirkovic2014class}. 

The training phases contain various difficulties, but all on the same topic. The decision component assigns exactly one task from the given phase. This assignment is based on the performance in previous phases and on the pre-training assessment. The performance is measured with time characteristics, used commands, submitted answers, and a solution taken in the phase. The tasks are denoted as T1, T2, \dots, T\emph{N}, where T1 represents the most difficult task in the phase and T\emph{N} the easiest. Further, the decision component processes the students' performance and knowledge to assign a~suitable task from the training phase.

Finally, the post-training questionnaire (Q) is an optional part of training, which enables instructors to collect immediate feedback from the participants.

\subsection{Model}
The decision component (P$_\text{D}$) is powered by a mathematical model, which assigns each student the most suitable task in each phase. The model uses binary vectors containing the performance metrics and a list of pre-configured weight matrices to set up the model. We use some of the performance metrics presented in the review of technical metrics for cybersecurity training~\cite{maennel2020learning}. 

\subsubsection*{Model Formulation}
Let us denote the following variables:\\
$\boldsymbol{p}$, $\boldsymbol{k}$, $\boldsymbol{a}$, $\boldsymbol{t}$, and $\boldsymbol{s}$ are the binary vectors on the correctness or incorrectness of prerequisites for a particular training phase. Vector $\boldsymbol{p}$ is defined as follows:
$\boldsymbol{p} = \begin{pmatrix}p_1 & p_2 & \dots & p_m\end{pmatrix}$, where $m$ is the number of rows. The other vectors use the analogous notation.
\begin{itemize}
    \item $\boldsymbol{p}$ represents the answers from the pre-training assessment,
    \item $\boldsymbol{k}$ indicates if the student used the expected key commands in the command line in the given task,
    \item $\boldsymbol{a}$ denotes whether the student used expected answers to the task,
    \item $\boldsymbol{t}$ contains the information if the task was completed in a~predefined time, and
    \item $\boldsymbol{s}$ contains the information whether the student asked to reveal the solution for the task,
    \item $\boldsymbol{W}$ is the matrix with weights for the individual phases' metrics.
\end{itemize} 

The model is defined by the \Crefrange{eq:weightMatrix}{eq:taskAssignment}.

By \Cref{eq:weightMatrix}, we get the \emph{weight matrix} that is specific for each training phase. The number of weight matrices is equal to the number of training phases. The weights represent the relationships between phases and their metrics. The value of the weight determines the importance of the metric to the phase. For instance, consider a training with six phases where the third phase deepens the topic exercised in the first phase. In this case, we set the weights in the third matrix so that the selected weights for the metrics from the first phase are non-zero. The other performance metrics with weights set to zero are ignored. The weights have to be manually set by the instructor since each training is unique. The symbols $\alpha,\beta,\gamma,\delta,\varepsilon$ denote the columns in the weight matrices and the $i=1,\dots, m$ are the rows in the weight matrices.

By \Cref{eq:achievedPerformance} we get the \emph{student's performance} based on the defined metrics and their weights for completed phases. The value of the performance is in the interval of $[0,1]$. In~\Cref{eq:achievedPerformance}, $s$ is multiplied by $a$, $k$, and $t$ to distinguish between students who satisfy $a$, $k$, and $t$ metrics without using a solution and solved the task on their own.

By \Cref{eq:taskAssignment} we get \emph{the number of the most suitable task} in phase $x$ for a particular student (1 is T1, 2 is T2, and so on).

\begin{align}\label{eq:weightMatrix}
\boldsymbol{W}^{(x)} = \left(w^{(x)}_{ij}\right), i=1,\dots,m,\:\:\: j=\alpha,\beta,\gamma,\delta,\varepsilon
\end{align}
\begin{equation}\label{eq:achievedPerformance}
        f(x)=\displaystyle\frac{\sum\limits_{i=1}^{x} \left[p_{i}w_{i\alpha}^{(x)} + s_{i} \left(k_{i}w_{i\beta}^{(x)} + a_{i}w_{i\gamma}^{(x)} +   t_{i}w_{i\delta}^{(x)} + w_{i\varepsilon}^{(x)}\right)\right]}
        {\sum\limits_{i=1}^{x} \left(w_{i\alpha}^{(x)}  + w_{i\beta}^{(x)} + w_{i\gamma}^{(x)} + w_{i\delta}^{(x)} + w_{i\varepsilon}^{(x)}\right)}
\end{equation}
\begin{align}\label{eq:taskAssignment}
    T_{x} = 
    \begin{cases} 
    n_x, \qquad\qquad\qquad\qquad\qquad\qquad \text{if } f(x) \text{ is equal to } 0\\
    \text{trunc}(n_x [1-f(x)])+1, \:\:\:\:\:\:\qquad \text{otherwise}
    \end{cases}
\end{align}

% description \lfloor a \rfloor is truncate
where:
\begin{align*}
    x= & \text{ the phase a student is entering},\\
    T_{x}= & \text{ the most suitable task of the phase $x$ for the student}, \\
    n_x= & \text{ the number of variant tasks in the phase $x$}, \\
    p_i= & 
    \begin{cases} 
        1, \: \text{if}\: \text{question group $i$ from A is correctly answered}\\
        0, \: \text{otherwise},
    \end{cases} \\
    k_i= & \text{ commands corresponding to the phase $i$ were used}, \\
    e_i= & \text{ expected time to complete of the phase $i$}, \\
    o_i= & \text{ student's completion time in the phase $i$}, \\
    t_i= &
    \begin{cases} 
        1, \: \text{if}\: o_i < e_i \text{ in phase $i$}\\
        0, \: \text{otherwise}, 
    \end{cases}\\
    s_i= &
    \begin{cases}
         1, \: \text{if}\: \text{the solution of the phase $i$ is \emph{not} displayed}\\
         0, \: \text{otherwise}, 
    \end{cases}\\
    a_i= & \text{ answers corresponding to the phase $i$ were submitted}.
\end{align*}

\subsubsection*{Model Assumptions}
The proposed model requires several assumptions that must be met by any system that would use it for hands-on cybersecurity training.
\begin{itemize}
    \item The learning environment has to collect the required data: commands typed by the students $k$, phase completion time $t$, the action of displaying the solution $s$, the submitted answers $a$, and the pre-training assessment answers $p$.
    \item The model expects that some tasks are related; otherwise, it will heavily rely only on the pre-training assessment that may not be sufficient to capture student's proficiency.
    \item The pre-training assessment question groups have to be mapped to the training phases to distinguish the level of knowledge and self-reported skills for a particular phase.
    \item The model assumes that the tasks in the phases are sorted so that the T1 is the most difficult task, T2, \dots, T\emph{N-1} are easier tasks than T1, and T\emph{N} is the easiest task.
\end{itemize}

To ease the unified design and run of the training, we add the following constraints that simplify the model assumptions:
\begin{itemize}
    \item The students' performance in a phase is evaluated in the same way in all tasks.
    \item The observed metrics are binary. Other metrics of students' performance, such as similarity of the submitted answers to the correct ones, are either unavailable or ignored.
\end{itemize}

The model was developed with the aim to reinforce the cybersecurity training with respect to the commonly used performance metrics \cite{maennel2020learning}. Nevertheless, it can be applied in any domain collecting such data.

\section{Case Study Setup}\label{sec:case_study_setup}
We describe the methods of the case study that uses the proposed adaptive training format and model. The case study uses data collected from 24 participants. The goal is to evaluate whether the proposed format and model are useful for adaptive hands-on cybersecurity training. In particular, we investigate whether the participants' experience is improved and if they successfully finish the training in a timely manner.

\begin{figure*}[!ht]
    \centering
    \includegraphics[scale=0.50]{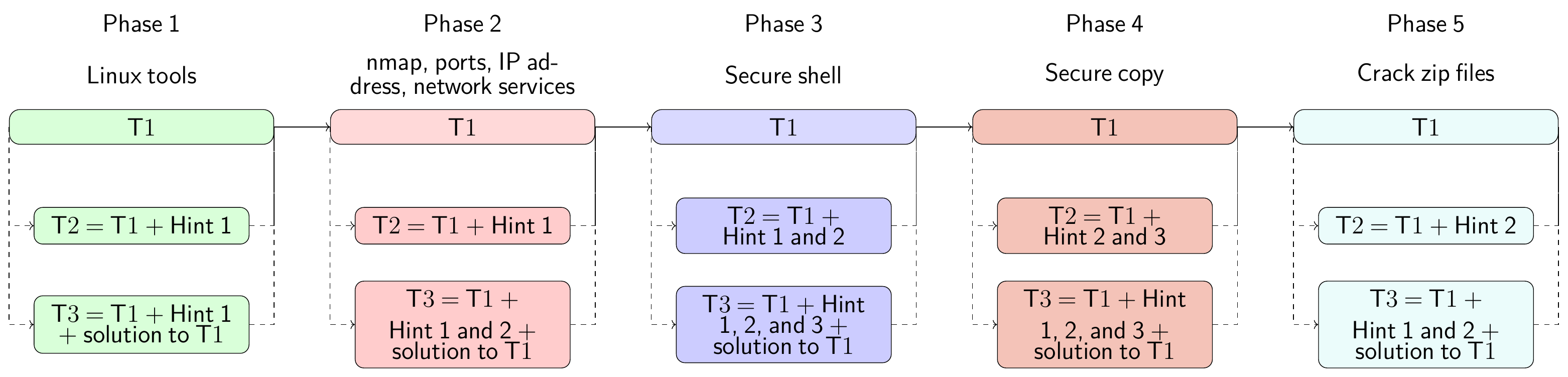}
    \caption{Phases of the adaptive training instance that follows the proposed generic format. Assignments of tasks contain assignments of base tasks and new or existing hints featured in base tasks.}
    \label{fig:adaptive_game_phases}
\end{figure*}

\subsection{Teaching Context and Participants}
The case study involved three training sessions held remotely in December 2020 and January 2021 at KYPO CRP~\cite{2021-FIE-kypo-csc}. 21 participants were undergraduate students of the Masaryk University, and three were graduates with one, two, and 12 years of professional experience in IT. All the participants provided informed consent to use the collected data for research purposes.

We designed a new adaptive training consisting of five interrelated phases. Each phase consists of tasks of various difficulty on the same topic. The phases and tasks were designed by one author and validated by the others. Then, the training was deployed to the KYPO CRP. At the time of the experiment, the learning environment did not provide the support for the proposed adaptive training format (presented in~\Cref{sec:design}). We implemented complementary software to process the data required by the model. The data were automatically collected and provided by the learning environment and manually entered into the complementary software by the authors after each phase.

At the beginning of the training session, students were asked to fill in the pre-training assessment and read the introduction of the training, including all necessary technical settings. Then, we assigned each student the most suitable task from the first training phase computed by the model. Once the student finished the training phase, they notified us, and we asked them to be patient while we entered the data into the complementary software. It calculated the suitable task in the next training phase (this corresponds to the P$_\text{D}$ nodes in~\Cref{fig:adaptive-game-structure-example}). Finally,~after finishing all the training phases, we asked the students to fill in the post-training questionnaire. After the training, all the data were anonymized so that they could not be attributed to a specific participant.

\subsection{Pre-training Assessment}
Given the limited time allocated to our training (one and half hours), we used a short pre-training self-assessment presented in~\Cref{tab:preGameAssessmentQuestions}. The self-assessment included the following question: \textit{What is your level of skill in the areas below?} and eight areas. The answer \textit{High} means you are able to complete the task very quickly and without much effort. \textit{Medium} means you are able to do it with standard effort. \textit{Low} means you have little experience with that. \textit{None} means you have no experience with that. We considered the students to have sufficient skills if they answered \textit{High} or \textit{Medium}.

\begin{table}[h!]
    \centering
    \caption{Wording of pre-training assessment questions and answers collected from 24 students. \faCheck~indicates a student sufficient skill and \faRemove~insufficient.}
    \label{tab:preGameAssessmentQuestions}
    \small
    \begin{tabular}{P{0.4cm}|p{3.7cm}|P{0.8cm}|P{0.9cm}|P{0.7cm}}
    \hline
    \textbf{No.} & \textbf{Question} & \textbf{Phase} & \multicolumn{2}{c}{\textbf{Answers}}\\  \hline
        & What is your level of skill in the areas below: & & &  \\ \hline
        Q1 & \texttt{msfconsole} interface         & none    & 2 \faCheck & 22 \faRemove  \\ 
        Q2 & WinSCP                           & none    & 5 \faCheck & 19 \faRemove  \\ 
        Q3 & build Java projects using Maven     & none    & 12 \faCheck & 12 \faRemove  \\
        Q4 & zip and unzip files in CLI             & 5       & 15 \faCheck & 9 \faRemove  \\
        Q5 & download and transfer files into the server & 4  & 14 \faCheck & 10 \faRemove  \\
        Q6 & connect to a server securely      & 3       & 16 \faCheck & 8 \faRemove  \\
        Q7 & search open ports               & 2       & 13 \faCheck & 11 \faRemove  \\
        Q8 & basic Linux commands             & 1       & 21 \faCheck & 3 \faRemove  \\
        \hline
    \end{tabular}
\end{table}

Questions Q4--Q8 were related to topics featured in our training. To avoid the disclosure of the phase topics by the wording of questions in the questionnaire, we added three distractor questions (Q1, Q2, and Q3) about topics not included in the training. The order of the questions differed from the order of the related training phases.

\subsection{Adaptive Training Phases}
The training in this study consists of five phases depicted in~\Cref{fig:adaptive_game_phases}. Each training phase features one base task and two variant tasks. Further, each phase features a task presenting the step-by-step solution. This was a last-resort task for students who would not match any phase prerequisites. In the first training phase, basic Linux tools are practiced in three variant tasks (T1, T2, and T3). Task T2 contains the same assignment as T1 and provides Hint~1. The third task T3 contains the assignment from T1 with Hint~1 and the solution to that task. The subsequent training phases apply the same pattern that differs only in the content of the tasks, hints, and solution provided. The tasks were assigned to each student by the proposed model. The settings of P$_\text{D}$ for each training phase are designed using the presented model settings.

%% the place of the figure 2 (training phase including hints)

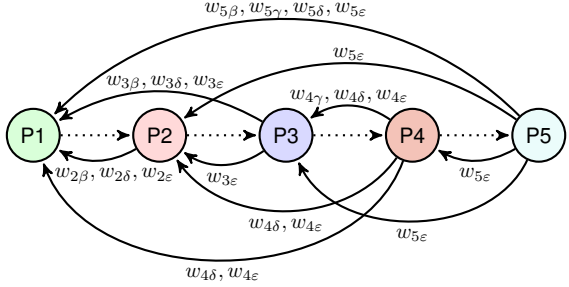
\begin{figure}[h!]
    \centering
    \begin{tikzpicture}[
        ->,                  % makes the edges directed
        >=stealth',          % makes the arrow heads bold
        shorten >=1pt,    
        auto,
        node distance=2.10cm,  % minimum distance between two nodes
        thick,
        scale=0.99,
        every node/.style={scale=0.8}, % oriģinally scale=0.8
        font=\sffamily
    ]
        \node[state,fill=green!15] (P1)        {P1};
        \node[state,fill=red!15] (P2) [right of=P1]      {P2};
        \node[state,fill=blue!15] (P3) [right of=P2]      {P3};
        \node[state,fill=anotherRed] (P4) [right of=P3]      {P4};
        \node[state,fill=phase5] (P5) [right of=P4]      {P5};
        
        % base paths
        \path 
        (P1) edge[dotted] node {} (P2)
        (P2) edge[dotted] node {} (P3)
        (P3) edge[dotted] node {} (P4)
        (P4) edge[dotted] node {} (P5)
        ;
        
        %% phase 2 predicates
        % commands, solution displayed
        \path (P2) edge [bend left=30] node[xshift=0.3cm] {$w_{2\beta},w_{2\delta},w_{2\varepsilon}$} (P1);
        
        %% phase 3 predicates
        % solution displayed
        \path (P3) edge [bend left=35] node[] {$w_{3\varepsilon}$} (P2);
        % commands, solution displayed, completed in time
        \path (P3) edge [bend right] node[yshift=0.4cm,xshift=0.1cm] {$w_{3\beta},w_{3\delta},w_{3\varepsilon}$} (P1);
    
        %% phase 4 predicates
        % solution displayed, completed in time, wrong flags
        \path (P4) edge [bend right=35] node[yshift=0.4cm] {$w_{4\gamma},w_{4\delta},w_{4\varepsilon}$} (P3);
        % solution displayed, completed in time
        \path (P4) edge [bend left=55] node {$w_{4\delta},w_{4\varepsilon}$} (P2);
        % solution displayed, completed in time
        \path (P4) edge [bend left=70] node {$w_{4\delta},w_{4\varepsilon}$} (P1);
        
        %% phase 5 predicates
        % solution displayed
        \path (P5) edge [bend left] node {$w_{5\varepsilon}$} (P4);
        \path (P5) edge [bend left=65] node {$w_{5\varepsilon}$} (P3);
        \path (P5) edge [bend right=35] node[yshift=0.4cm] {$w_{5\varepsilon}$} (P2);
        % commands, solution displayed, completed in time, wrong flags
        \path (P5) edge [bend right=45] node[yshift=0.4cm] {$w_{5\beta},w_{5\gamma},w_{5\delta},w_{5\varepsilon}$} (P1);
    \end{tikzpicture}
    \caption{The relationships between all training phases. P\emph{X} is a phase $x$ and $w_{xy}$ is weight for phase $x$ and metric $y$.}
    \label{fig:adaptive-game-relationships}
\end{figure}

\subsection{Model Settings}\label{sec:modelSettings}
To use the model, we must set the weights in the weight matrix $\boldsymbol{W}$ for each training phase, see \Cref{eq:weightMatrix}. These weights indicate the relationships between training phases. For simplicity, we set these weights to zero or one in our case study. One indicates the relationship and zero indicates that there is no relationship between the phases. Each training phase is related to a particular question group from the pre-training assessment. The relationships between training phases in our training are shown in~\Cref{fig:adaptive-game-relationships}.

% the place of figure (fig:adaptive-game-relationships)

\section{Results and Discussion}\label{sec:results}
In this section, we report the results of the study and summarize our experience with adaptive learning in cybersecurity hands-on training.

\subsection{Adaptive Training Results}\label{sec:adaptiveTrainingResults}
\begin{figure}[h!]
    \centering
    \includegraphics[width=\columnwidth]{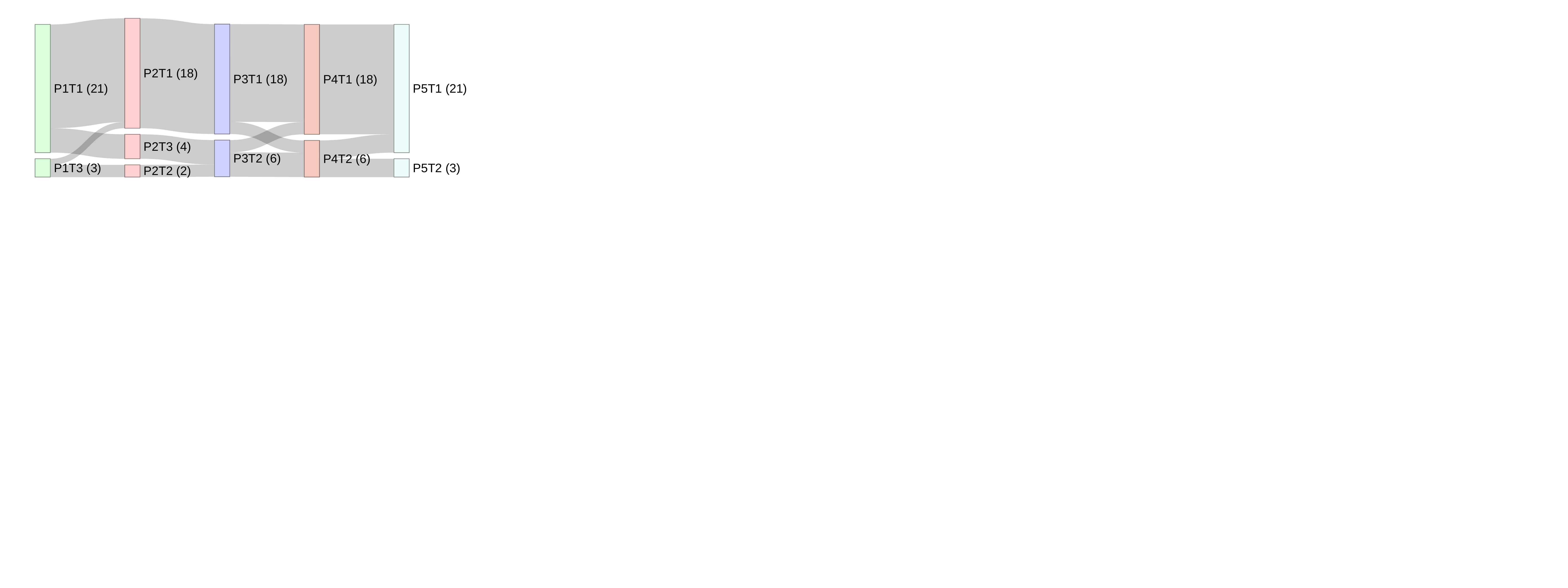}
    \caption{Transitions of 24 participants between particular tasks in training. P\emph{X}T\emph{Y} denotes task T\emph{Y} in the phase P\emph{X}. The number of participants solving the task is in brackets.}
    \label{fig:adaptivephasesFlow}
\end{figure}

Using the ITS terminology, our case study examined \textit{student model} (the participants' performance), \textit{domain model} (the developed training and its phases with tasks), and the \textit{tutor model} (newly proposed model for assigning the most suitable tasks to each participant). \Cref{fig:adaptivephasesFlow} shows the transitions of 24 participants between tasks (P\emph{X}T\emph{Y}) in all training phases. We see that the participants went through different tasks in the training phases, which suggests that the participants' proficiency did not always match the base tasks. We believe this is natural, and the main reason why some participants failed to successfully complete the training in our previous hands-on training sessions.

The selection of tasks in the first training phase was based on answers from the pre-training assessment because no other performance metrics had been available yet. The three participants claimed that they were not familiar with the Linux operating system, so they played the easiest task in the first phase (P1T3). In the second phase, not only the answers from the pre-training assessment but also the participants' performance from the previous phase were available. The diversity of assignments of tasks to participants increased; the easier tasks were solved by six participants in total. The six participants did not complete the first phase in the expected time ($e_i$), two used too many commands ($k_i$), one displayed the solution ($s_i$), and 11 did not have experience with the tool required for phase two ($p_{i}$). It is evident that the participants face different issues during and after the first phase. That confirms our assumption that it is difficult to design static hands-on training suitable for all participants.

In the rest of the training phases, the model assigned the variant tasks to some participants because they were unable to complete the previous phases on time, exceeded the number of expected key commands (set to 10), or scored low in the pre-training assessment. Overall, even in this relatively small sample of participants, their paths through the training differ substantially. The worst performing participant received mostly the easiest tasks (P1T3, P2T2, P3T2, P4T2, and P5T2) and finished the training in 89 minutes, while the best performing participant completed the most difficult (base) tasks in 13 minutes. Regarding the successful completion of the training, 88\% of participants successfully completed the training without any solution taken.

\subsection{Post-training Questionnaire}\label{sec:postTrainingQuestionnaire}
\begin{figure}[h!]
    \centering
    % boxplotdata.dat form -> index median box_top box_bottom whisker_top whisker_bottom
    \scalebox{0.80}{ % \scalebox{<h-scale>}[<v-scale>]{<content>} 
   %  \small
        \begin{tikzpicture}%[fixed ratio=16:9]
            \begin{axis} [
                xtick={1,2,3,4,5,6},
                xticklabels={Q1, Q2, Q3, Q4, Q5, Q6},
                ytick={1,2,3,4,5},
                yticklabels={Not at all, Slightly, Moderately, Much, Very much},
            ]
                \addplot [box plot median] table [opacity=0.5]{boxplotdata.dat};
                \addplot [box plot box] table {boxplotdata.dat};
                \addplot [box plot top whisker] table {boxplotdata.dat};
                \addplot [box plot bottom whisker] table {boxplotdata.dat};
            \end{axis}
        \end{tikzpicture}
    }
    \caption{Answers gathered in post-training questionnaire (n = 24).}
    \label{fig:postGameAssessment}
\end{figure}
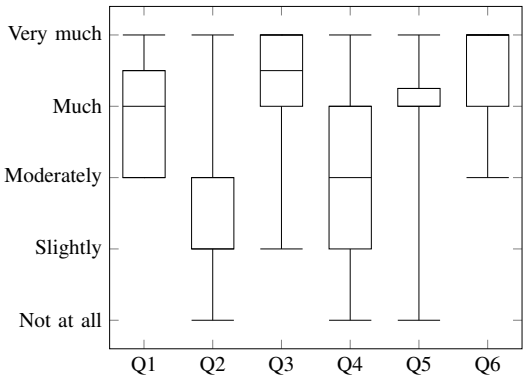

Immediately after the training session, we asked the participants for their feedback in the online survey. \Cref{tab:postGameAssessmentQuestions} lists the questions (Q1--Q6) and \Cref{fig:postGameAssessment} summarizes the answers.

\begin{table}[h!]
    \centering
    \caption{Wording of the post-training questionnaire.}
    \label{tab:postGameAssessmentQuestions}
    \small
    \begin{tabular}{p{0.3cm}|p{7.3cm}}
    \hline 
    \textbf{No.} & \textbf{Question} \\  \hline
        Q1 & Did you feel the tasks were designed so that you can complete the training in a timely manner?  \\ 
        Q2 & Did you feel you got stuck at some point during the training?  \\ 
        Q3 & How much did you enjoy the training?  \\
        Q4 & Did you feel the training should be more difficult for you?  \\
        Q5 & Did you feel you would like the training to be longer with additional tasks to solve?  \\
        Q6 & Would you like to play more cybersecurity training sessions like this one?  \\
        \hline
    \end{tabular}
\end{table}

In the first question (Q1), the participants reported that the tasks were appropriately designed so that they have successfully completed the training in time. This question has an additional option \textit{Did not receive any}, which was chosen by nine participants. The second question (Q2) was crucial. Five participants reported \textit{Not at all}, eight participants \textit{Slightly}, seven participants \textit{Moderately}, three participants \textit{Much}, and only one participant reported \textit{Very much}. This suggests the training session went relatively smoothly and the majority of participants did not experience anything that would lead them to frustration or premature training termination. In the fourth question (Q4), only one participant reported that the training should be more difficult. This suggests the need for designing more difficult tasks for the very experienced participants who may get bored if the base tasks are too easy for them. Answers to Q5 indicate that the participants engaged in the training and would like to continue if it would be possible. To conclude, the participants reported (Q3 and Q6) that they enjoyed the training and that they would like to join another adaptive hands-on training in cybersecurity.

\subsection{Limitations}\label{sec:expLessonsLearnedLimitations}
In contrast to other fields, cybersecurity hands-on trainings are usually held in a group of lower tens of participants. Therefore, we believe 24 is a sufficient number of participants to evaluate the created adaptive training format using the newly developed model. 

Given the limited time allocated to our training (one and half hours), we used a short pre-training self-assessment. Nevertheless, for training sessions with a larger time allocation, we recommend adding questionnaire quizzes along with self-reported skills \cite{vsvabensky2018challenges,mirkovic2014class}.

Although the model is not limited to a specific design of variant tasks, we created the variant tasks by changing the text of the assignment (by uncovering particular steps or providing hints). Another option would be to modify the environment (i.e., network and hosts) for the variant tasks. That would give us more freedom in creating the variant tasks.

The model allows including an arbitrary number of tasks in each phase. In our study, we designed three tasks for each phase. Providing more tasks may increase the probability that the participant will get a more suitable task. However, designing more tasks increases instructors' effort to prepare the training.

\section{Conclusions}
\label{sec:conclusion}
Hands-on cybersecurity training sessions usually use static scenarios with limited or no adaptiveness. In this paper, we analyzed student performance and failures in the past training sessions. This led us to propose a new adaptive training format using a graph structure and a generic tutor model. The tutor model is used to assign the most suitable task to each student in each training phase.

Using this innovation, we try to assign the students the optimal path through the training so that they learn as much as possible and keep being motivated for further learning. For these purposes, we developed a new adaptive training format and held three training sessions with 24 participants in total. The results showed that adaptive learning can increase the students' ability to successfully complete the hands-on training, and thus increase the positive students' experience. Further, it showed that the proposed tutor model is useful and can be used for various training sessions with different topics. The students mostly reported that they did not get stuck in any phase of the training and that they enjoyed the training.

To ease the adoption of the proposed innovation, we publish data from the training sessions, together with the model, at~\cite{seda2021datasetreinforcing}.

Finally, we provide recommendations for instructors developing adaptive training and ideas for future work.

\subsection{Recommendations for Instructors}
To effectively run the adaptive training using the proposed training format and model, consider the following recommendations.

\paragraph{The pre-training assessment questionnaire should be simple and brief}{ Cybersecurity education sessions are usually held in a limited time frame. The questionnaire should not consume a large amount of that time, but must still follow best practices for educational assessment~\cite{astin2012assessment, petty2009}. For example, explain the importance of the questionnaire clearly and explicitly to students.}

\paragraph{Adjust the weights in the model carefully}{Setting weights in the weight matrices determines the relationships between individual phases and their metrics. Based on that, participant performance for the given phase is calculated. If weights are adjusted incorrectly, the student can get an inappropriate task and may get bored or stuck in the phase.}

\paragraph{Design at least three tasks for each phase}{Without enough tasks, the model cannot assign a suitable task for differently performing participants. The base task should be as difficult as possible to target the most experienced participants, and one of the variant tasks should be as easy as possible (step-by-step solution) to encourage less experienced participants.}

\paragraph{Allocate more time for students to complete the base phases than you expect}{Since assignments of the base tasks are intentionally vague to allow exploring the phase topic, students need enough time for some trial and error. However, our experience shows that the majority of instructors estimate too short time to complete.}

\subsection{Future Work}
We proposed a generic model and set its parameters for a particular training session. Therefore, future work should investigate more model metrics and advanced parameter settings.
The model decides to move up or down in difficulty for students. But, for example, a student knowing a topic may need a refresher; or a student not knowing a topic may need the challenge to awaken their interest. Further, in our case study, we designed three tasks for each phase and we did not study the effect of a different number of tasks. These issues will be addressed in our future work.

Finally, the decision component (P$_\text{D}$) was provided by the complementary software that required us to do some analytical tasks manually. In our future work, we will enhance this component to be fully automated and integrate it with the KYPO CRP to fully support the proposed adaptive training format. 

\section*{Acknowledgment}
This research was supported by the Security Research Programme of the Czech Republic 2015--2022 (BV III/1--VS) granted by the Ministry of the Interior of the Czech Republic under No. VI20202022158 -- Research of New Technologies to Increase the Capabilities of Cybersecurity Experts.

\balance

\bibliographystyle{IEEEtran}
\bibliography{references}

\end{document}